%% file: alvaro-hbase-qod-cloudcom2013.tex
\documentclass[10pt, conference, compsocconf]{IEEEtran}

\pdfoutput=1

\usepackage{epstopdf}
\usepackage[latin1]{inputenc}
\usepackage[bookmarks=false]{hyperref}
\usepackage[T1]{fontenc}
\usepackage{mathtools}
\usepackage[pdftex]{graphicx}
\usepackage{algorithmic}
\usepackage{algorithm}
\usepackage{multicol}
\usepackage[justification=centering]{caption}
\usepackage{float}
\usepackage{color}
\graphicspath{ {.} }

\begin{document}
\bibliographystyle{IEEEtran}

\title{Quality-of-Data for Consistency Levels in Geo-replicated Cloud Data Stores}

\author{\IEEEauthorblockN{\'{A}lvaro Garc\'{i}a-Recuero}
\IEEEauthorblockA{Instituto Superior T\'{e}cnico (IST)\\
INESC-ID Lisboa / Royal Institute of Technology (KTH)\\
Lisbon, Portugal / Stockholm, Sweden\\
alvaro.recuero@ist.utl.pt}
\and
\IEEEauthorblockN{S\'{e}rgio Esteves}
\IEEEauthorblockA{Instituto Superior T\'{e}cnico (IST)\\
INESC-ID Lisboa\\
Lisbon, Portugal\\
sesteves@gsd.inesc-id.pt}
\and
\IEEEauthorblockN{Lu\'{i}s Veiga}
\IEEEauthorblockA{Instituto Superior T\'{e}cnico (IST)\\
INESC-ID Lisboa\\
Lisbon, Portugal\\
luis.veiga@inesc-id.pt}
}

\maketitle
\begin{abstract}
\input{abstract}
\end{abstract}

\begin{IEEEkeywords}
Geo-replication; HBase; NoSQL; YCSB;
\end{IEEEkeywords}

\thispagestyle{empty}
\section{Introduction}
\input{introduction}

\section{Background and Related Work}
\input{relatedwork}

\section{HBase-QoD Architecture}
\input{architecture}

\section{Implementation Details}
\input{implementation}

\section{Simulation and Evaluation}
\input{evaluation}

\section{Conclusion}
\input{conclusion}

\section*{Acknowledgments}
\small{This work was partially supported by national funds through FCT -- Funda\c {c}\~{a}o para a Ci\^{e}ncia e a Tecnologia, projects PTDC/EIA-EIA/113613/2009, PTDC/EIA-EIA/108963/2008, PEst-OE/EEI/LA0021/2013.}

\bibliography{bibliography}
\end{document}

%% file: abstract.tex
Cloud computing has recently emerged as a key technology to provide individuals and companies with access to remote computing and storage infrastructures. In order to achieve highly-available yet high-performing services, cloud data stores rely on data replication. However, providing replication brings with it the issue of consistency. Given that data are replicated in multiple geographically distributed data centers, and to meet the increasing requirements of distributed applications, many cloud data stores adopt eventual consistency and therefore allow to run data intensive operations under low latency. This comes at the cost of data staleness. In this paper, we prioritize data replication based on a set of flexible data semantics that can best suit all types of Big Data applications, avoiding overloading both network and systems during large periods of disconnection or partitions in the network. Therefore we integrated these data semantics into the core architecture of a well-known NoSQL data store (\emph{e.g.}, HBase), which leverages a three-dimensional vector-field model (regarding timeliness, number of pending updates and divergence bounds) to provision data selectively in an on-demand fashion to applications. This enhances the former consistency model by providing a number of required levels of consistency to different applications such as, social networks or e-commerce sites, where priority of updates also differ. In addition, our implementation of the model into HBase allows updates to be tagged and grouped atomically in logical batches, akin to transactions, ensuring atomic changes and correctness of updates as they are propagated.

%% file: introduction.tex
In distributed systems in general and Cloud Computing specifically, data replication is becoming a major challenge with large amounts of information requiring consistency and high availability as well as resilience to failures. There have been several solutions to that problem, none of them applicable in every case, as they all depend of the type of system and its end goals. As the CAP theorem states~\cite{Brewer:2002}, one can not ensure the three properties of a distributed system all at once, therefore applications usually need to compromise and choose two out of three between consistency, availability and tolerance to partitions in the network.

Nowadays, in many data center environments is key to understand how one makes such distributed systems scalable while still delivering good performance to applications. Data availability is for instance always a desired property, while a sufficiently strict level of consistency should be used to handle data effectively across locations without long network delays (latency) and optimizing bandwidth usage.

There are a number of existing systems where data semantics are analyzed to provide operations with faster (eventual) or slower (stronger) consistency without compromising performance~\cite{Li:2012}. In some, causal serialization and therefore commutative updates are provided also based on data semantics, but require redesigning application data types~\cite{Saphiro:2011} or intercepting and reflecting APIs via middleware~\cite{Vfc3:2012}. Unlike linearizability, eventual consistency does work well for systems with shared distributed data to be queried and/or updated, because updates can be performed on any replicas at any given time~\cite{Burckhardt:2012}. It is then easier to achieve lower latency, so most systems implement eventual consistency in order to avoid expensive synchronous operations across wide area networks and still keeping data consistent through low latency operations in large geo-located deployments.

HBase is a well-known and deployed open source cloud data store written and inspired on the idea of BigTable~\cite{Carstoiu:2010} which targets the management of large amounts of information. HBase does not provide strong consistency outside of the local cluster itself. Eventuality is the promise and a write-ahead log maintained for that.

This work introduces \emph{HBase-QoD}, a replication module that integrates into HBase mechanisms targeting applications which might also require finer-grain levels of consistency. Other systems use a variant of Snapshot Isolation techniques~\cite{Sovran:2011}, which works within, but not across data centers. Others, like the conit model, are based on generality but not practicality~\cite{Duke:2001}. We find the later to be more rewarding to users in terms of quality of data within a fully functional and reliable data storage system, achieving optimization of resources during geo-replication and consequently significant cost savings. We propose an architecture with custom levels of consistency, providing finer-grain replication guarantees through data semantics. Application behavior can be therefore turn into a key and more efficient shift into the consistency paradigm. This is reflected in this paper by modifying and extending eventual consistency, with an innovative approach used to tune its replication mechanisms, originally developed for treating all updates equally.

\subsection{Contributions}
The main contributions are focused on what other consistency properties HBase can provide between different geo-located clusters at the replication level, that is, using a flexible and tunable framework, developed for treating groups of updates tagged for replication in a self-contained manner. The work presented includes the design and development of the model to be applied to non-relational cloud-based data stores.

We take a step forward from the eventual consistency model at inter-site replication scenarios with HBase deployments to prove the validity of the model. It is possible to evaluate it by using an adaptive consistency model that can provide different levels of consistency depending of the desired Service Level Objective (SLO) for the Quality-Of-Data (QoD) fulfillment. The idea of QoD fulfillment is based on the percentage of updates that need to be replicated within a given period using a three-dimensional vector model \emph{K}.

We also propose to extend HBase client libraries in order to provide grouping of operations during replication, where each of the groups can provide the level of consistency required: ANY, IMMEDIATE, or even with a specific custom bound. To achieve this we modify HBase libraries (Htable). Grouping of operations occurs from the source location before replication actually occurs, so apart from the multi-row atomically defined model in HBase, a more versatile system can also provide atomically replicated updates beyond the row-level (e.g., column families or combinations of the fields in a row in HBase). The work is also an informal contribution that we aim to turn into a formal one to complete the efforts of a "pluggable replication framework" as proposed by the Apache HBase community.~\footnote{Priority queue sorted replication policy, A. Purtell, August, 2011.}

\subsection{Roadmap}
In the next sections of the paper we have a brief overview of fundamental consistency models and background work in this area of distributed systems, having special focus on the concept of eventual versus strong consistency and what possible variations of the two can exist in the middle of the spectrum. As intermediate approach, we devise HBase-QoD, a HBase extension to apply QoD defined through a three-dimensional vector-field model inspired on~\cite{Santos:2010}. Data semantics are defined and enforced with a vector \emph{K} ($\theta$, $\sigma$, $\nu$), representing Time, Sequence and Value respectively.

The rest of the paper is organized as follows, related work in Section 2, our HBase extension architecture in Section 3, the implementation details in Section 4, and evaluation in Section 5. The evaluation results show that from the architectural point of view our solution integrates well in HBase and provides the corresponding vector-bounded replication guarantees. Finally, with Section 6 we conclude this work.

%% file: relatedwork.tex
HBase is open source, and its architecture is based in previous work at Google, BigTable~\cite{Chang:2006}, a distributed, persistent and multi-dimensional sorted map. HBase is being used for instance at Facebook data-centers for structured storage of the messaging and user data in partial replacement of Cassandra~\cite{Lakshman:2010}. Cassandra offers replica-set consistency tuning, but not divergence bounded consistency regarding data semantics. In geo-located and replicated remote clusters, the system provides eventual guarantees to data consistency through RPC (Remote Procedure Call) mechanisms.

Eventual consistency might be sufficient in most cases. Although, complex applications require stronger consistency guarantees and can be difficult to manage. Due to that, there have been recent research efforts to address these shortcomings in geo-replicated data centers, with Google developing earlier in 2012 an evolution of BigTable that provides external consistency through atomic clocks for instance, Spanner~\cite{Corbett:2012}. This makes applications highly-available while ensuring as much synchronicity among distant replicas as possible and more importantly, atomic schema changes. Data locality is also an important feature for partitioning of data across multiple sites. Spanner does use Paxos for strong guarantees on replicas.

Strong consistency does not work well for systems where we need to achieve low latency. So the reason for most systems to use eventual consistency is mostly to avoid expensive synchronous operations across wide area networks. In other cases such as COPS~\cite{Lloyd:2011} causality is guaranteed, although it does not guarantee the quality of data by bounding divergence, which can still lead to outdated values being read. Previous inspiring work from \cite{Yu:2000} also shows divergence bounding approaches to be feasible in that regard.

Systems such as PNUTS from Yahoo~\cite{Cooper:2008}, introduced a novel approach for consistency on a per-record basis. Therefore, it became possible to provide low latency during heavy replication operations for large web scale applications. As in our work, they provide finer-grain guarantees for certain data, so in other words, new updates are not always seen right away by the clients (which is the case also with our HBase extension), but only if strictly necessary. Keeping that in mind, it is not mandatory for applications to be highly-available and consistent both at once. That is applicable to our use case. Yahoo made the case for eventual consistency not being enough, and as in the case of social networks, stale replicas can introduce privacy issues if not handled adequately. We propose using operation grouping to resolve the consistency issue among blocks of updates more efficiently and in an straight forward manner by using several QoD levels.

%% file: architecture.tex
HBase-QoD allows for entries to be evaluated prior to replication, and it is based in the aforementioned vector-field consistency model, allowing for a combination of one or several of the parameters in K ($\theta$, $\sigma$, $\nu$), corresponding to Time, Sequence, and Value divergence in this case. Secondly, updates that collide with previous ones (same keys but different values) can also be checked for number of pending updates or value divergence from previously replicated items and, if required, shipped or kept on hold accordingly. The time constraint can be always validated every X seconds, and the other two through Alg.~\ref{algo1} as updates arrive. For the work presented here we use Sequence ($\sigma$) as the main vector-field constraint to showcase the model in practice. For this, we define a set of customized data structures, which hold the values of database row items due to be checked for replication on a particular data container. Containers are identified as \emph{tableName:columnFamily} for us, and can also be a combination of relevant row-fields from the data store schema with the purpose of differentiating data and apply semantics onto it.

In order to track and compare QoD fields (which act as constraints during replication) against stored updates, data \emph{containers} are defined for the purpose, controlling both current and maximum (then reset) bound values. Therefore, a QoD percentage is on the updates due to be propagated at once (e.g., using $\sigma$). This process is partly automated at the moment, with us just defining it at run-time (or by the developer later) adding a parameter into the HBase console that selects the desired vector-field bound percentage.

\begin{figure}[h]
\centering
\includegraphics[width=1.0\linewidth]{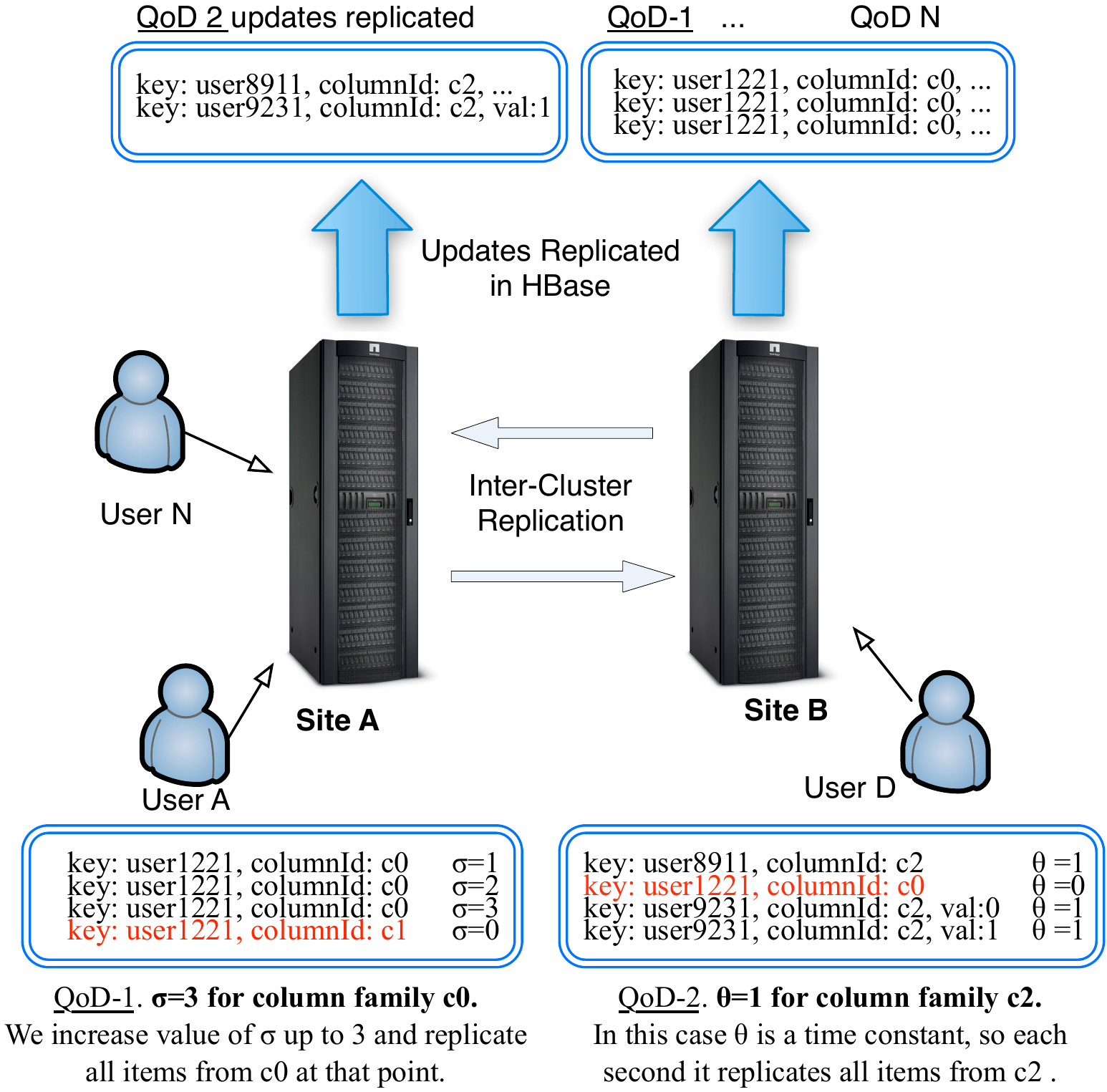}
\caption{HBase QoD high-level}
\label{fig-high-level}
\end{figure}

\paragraph*{Extensions to the HBase internal mechanisms}
A high-level view of the mechanisms introduced with HBase-QoD is outlined in Figure~\ref{fig-high-level}, and it is based in each case in an specific QoD bound applied to each defined data container per user. Replicating when the QoD is reached means in the case of QoD-1, sending $\sigma$ updates, 3, from User A (assume is not same data-container value) and each second for the User D with QoD-2 of $\theta$ (time, reset back to zero if reached) field, in this case also showing the \emph{last-writer wins} behavior on the remote side, user N, for a same propagating remote data-container value during replication. This overall architecture layout is presented in order to showcase a scenario where to rule updates selectively during geo-replication.

The original HBase architecture has built-in properties derived from the underlying HDFS layer. As part of it, the WALEdit data structure is used to store data temporarily before being replicated, useful to copy data between several HBase locations. The QoD algorithm (shown in Algorithm.~\ref{algo1}) uses that data structure, although we extend it to contain more meaningful information that help us in the management of the outgoing updates marked for replication. We extend HBase, handling updates due to be replicated in a priority queue according to the QoD specified for each of their data containers. Thereafter once the specified QoD threshold is reached the working thread in HBase, in the form of Remote Procedure Call, collects and ships all of them at once.

\paragraph*{Typical distributed and replicated deployment}
In distributed clusters Facebook is currently using HBase to manage the messaging information across data centers. That is because of the simplicity of consistency model, as well as the ability of HBase to handle both a short set of volatile data and ever-growing data, that rarely gets accessed more than once.

With the eventual consistency enforcement provided, updates and insertions are propagated asynchronously between clusters so Zookeeper is used for storing their positions in log files that hold the next log entry to be shipped in HBase. To ensure cyclic replication (master to master) and prevent from copying same data back to the source, a sink location with remote procedure calls invoked is already into place with HBase. Therefore if we can control the edits to be shipped, we can also decide what is replicated, when or in other words, how soon or often.

\begin{algorithm}
\caption{QoD algorithm for selecting updates using $\sigma$ criteria (with time
and value would be the same or similar) Returns true means replicate.}\label{algo1}
\label{alg1}
\begin{algorithmic}[1]
\REQUIRE $containerId$
\ENSURE $maxBound \neq 0$ and $controlBound \neq 0$
\WHILE{$enforceQoD (containerId)$}
\IF{$getMaxK(containerId) = 0$}
\RETURN $true$
\ELSE[$getactualK(containerId)$]
\STATE $actualK(\sigma) \leftarrow actualK(\sigma)+1$
	\IF{$actualK(\sigma) \geq containerMaxK(\sigma)$}
	\STATE $actualK(\sigma) \leftarrow 0$
	\RETURN $true$
	\ELSE
	\RETURN $false$
	\ENDIF
\ENDIF
\ENDWHILE
\end{algorithmic}
\end{algorithm}

\paragraph*{Operation Grouping}
\input{operation-grouping} 

%% file: operation-grouping.tex
At the application level, it may be useful for HBase clients to enforce the same consistency level on groups of operations despite affected data containers having different QoD bounds associated. In other words, there may be specific situations where write operations need to be grouped so that they can be all handled at the same consistency level and propagated atomically to slave clusters. 

For example, publication of user statuses in social networks is usually handled at eventual consistency, but if they refer to new friends being added (e.g., an update to the data container holding the friends of a user), they should be handled at a stronger consistency level to ensure they are atomically visible along with the list of friends of the user in respect to the semantics we describe here.

In order to not violate QoD bounds and maintain consistency guarantees, all data containers of operations being grouped must be propagated either immediately after the block execution, or when any of the QoD bounds associated to the operations has been reached. When a block is triggered for replication, all respective QoD bounds are naturally reset. 

To enable this behavior we propose extending the HBase client libraries to provide atomically consistent blocks.
Namely, adding two new methods to HTable class in order to delimit the consistency blocks: \textit{startConsistentBlock} and \textit{endConsistentBlock}. Each block, through the method \textit{startConsistentBlock}, can be parameterized with one of the two options: i) \textit{IMMEDIATE}, which enforces stronger consistency for the whole block of operations within itself; and ii) \textit{ANY}, which replicates groups of updates as a whole and as soon as the most stringent (smaller) QoD vector-field bound, associated with an operation inside the block, is reached.



%
%
%
%

%% file: implementation.tex
HBase replication mechanism is based in a Write Ahead Log (WAL), which must be enabled in order to be able to replicate between distant data centers. The process of replication is currently carried out asynchronously, so there is no introduced latency in the master server. Although, since that process is not strongly consistent, in write heavy applications a slave could still have stale data for an order of more than just seconds, and just until the last updates commit to local disk.

In our implementation we overcome the pitfalls of such an approach, and propose a QoD-vector to handle selected updates, thus lower values of it (e.g maxBound of $\sigma$ in the three dimensional vector K) enforce their delivery at the remote cluster earlier. For write intensive applications, that can be both beneficial in terms of reducing the maximum bandwidth peak-usage, while still delivering data according to application needs and with improved semantics.

\begin{figure}
\centering
\includegraphics[scale=0.6]{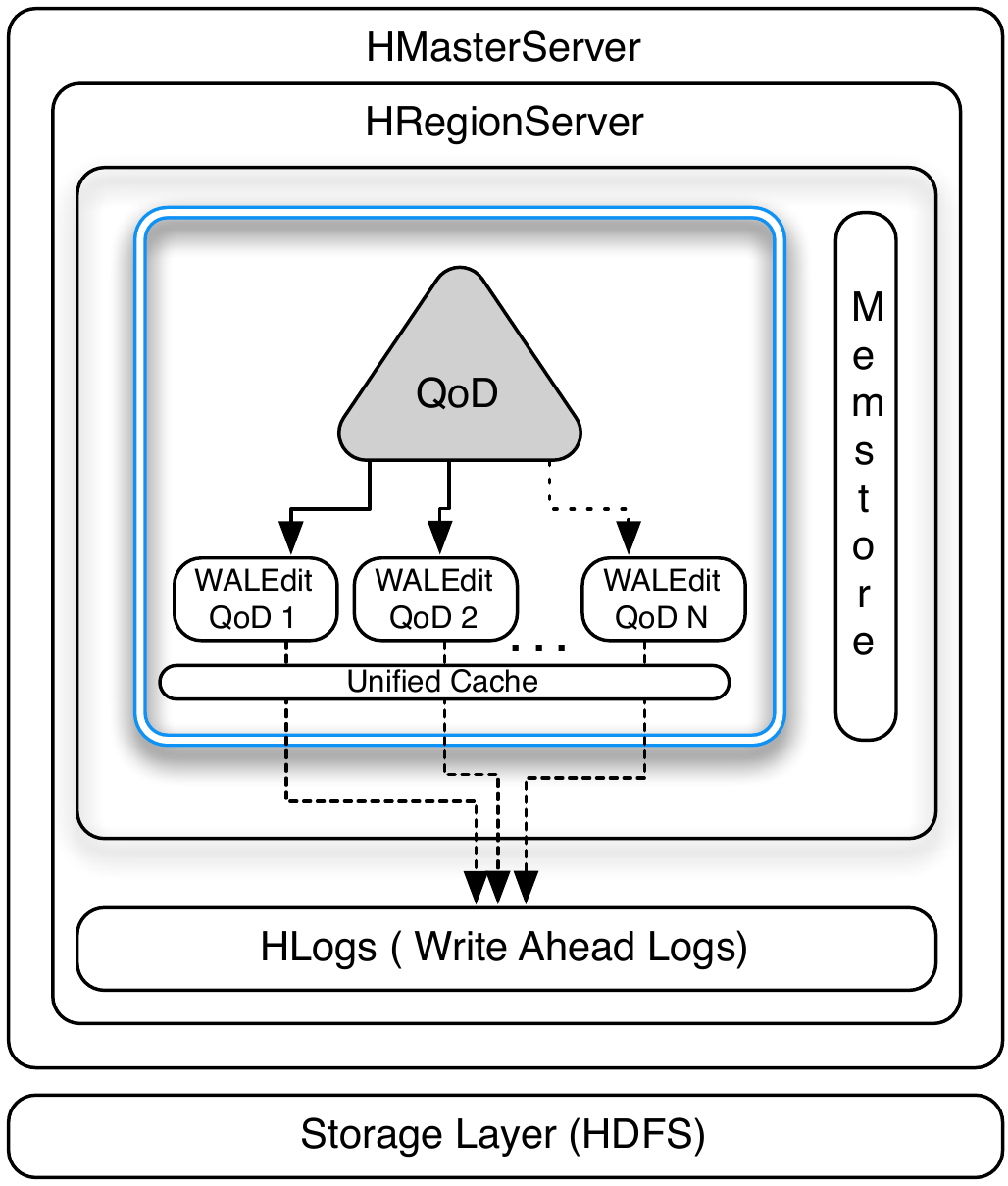}
\caption{HBase QoD operation}
\label{fig-uni-site}
\end{figure}
The QoD module in Figure~\ref{fig-uni-site} shows the implementation details introduced with HBase-QoD. We observe the module integrates into the core of the chosen cloud data store (HBase), intercepting incoming updates, and processing them into a priority queue named \emph{Unified Cache}, which is defined to store those for later replication to a given slave cluster.

Changing the logic of the shipping of edits in the write-ahead log, this process is therefore performed now according to the data semantics we define. Several data structures are required, some of them existing in HBase, as the WALEdit. That is in order to access different data containers that we later query to determine where and how to apply a given QoD at the row level (e.g. tablename:columnFamily). The data is replicated once we check the conditions shown in Algorithm 1 are met, and replication is triggered if there is a match for any of the vector-constraints (e.g $\sigma$). The use of the QoD is also applicable to the selection of those updates to be replicated according to a combination of any the three-dimensional vector constraints, not only $\sigma$. 

%% file: evaluation.tex
It has been already verified and presented in other reports and projects in the area of Hadoop, that a statically defined replication level is in itself a limitation, which therefore must be addressed and more efficiently adjusted in order to keep up with the scheduling of tasks. That is also related to the work here covered within HBase, as HDFS is its storage layer. A workaround on static replication constraints in HDFS and HBase is offering and enforcing on-demand replication with HBase-QoD and its vector-field model. During evaluation of the model, a test-bed of several HBase clusters has been deployed, having some of them using the HBase-QoD engine enabled for quality of data between replicas, and others running a regular implementation of HBase 0.94.8. All tests were conducted using 6 machines with an Intel Core i7-2600K CPU at 3.40GHz, 11926MB of available RAM memory, and HDD 7200RPM SATA 6Gb/s 32MB cache, connected by 1 Gigabit LAN.

\begin{figure}[b]
\centering
\includegraphics[width=1.0\linewidth]{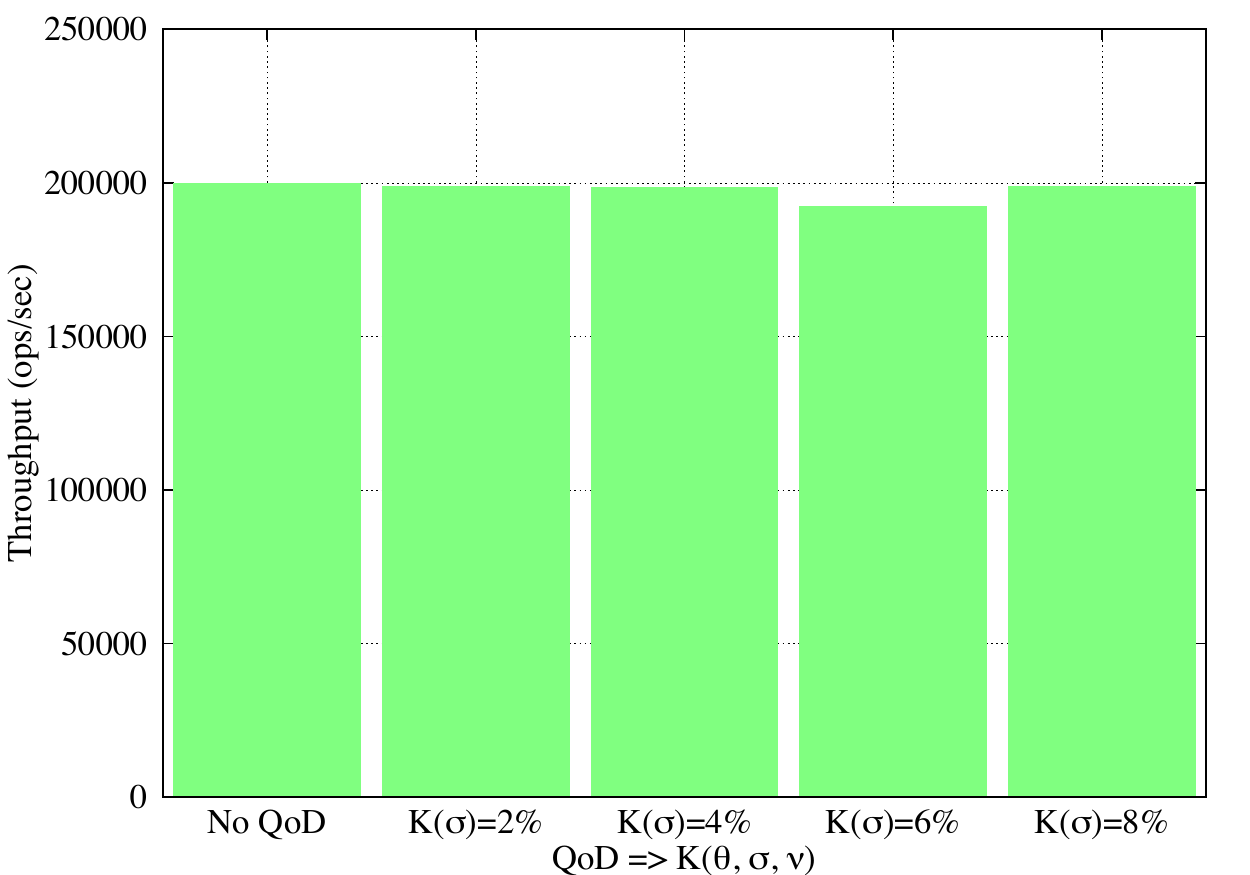}
\caption{Throughput for several QoD configurations}
\label{fig-throughput}
\end{figure}
We confirm the QoD does not hurt performance as we observe from the throughput achieved for the several levels of QoD chosen during the evaluation of the benchmark for our modified version with HBase-QoD enabled, Figure~\ref{fig-throughput}. The differences in throughput are irrelevant and mostly due to noise in the network, that is the conclusion after obtaining similar results to that one in several rounds of tests with the same input workload on the data store.

\begin{figure}[h]
\centering
\includegraphics[width=1.0\linewidth]{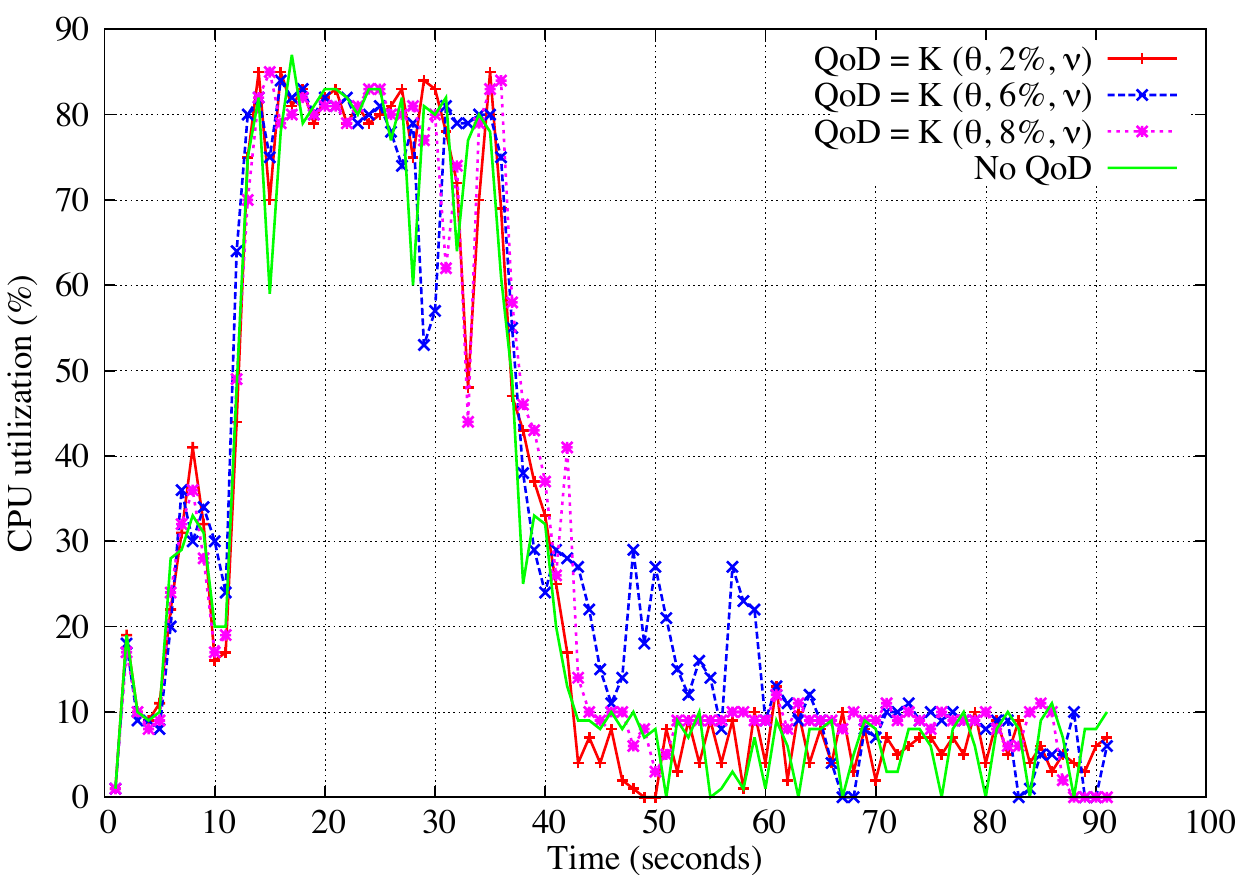}
\caption{CPU usage over time with QoD enabled}
\label{fig-cpu}
\end{figure}

Next we conducted as shown in Figure~\ref{fig-cpu}, and \emph{dstat} presents, an experiment to monitor the CPU usage using HBase-QoD. CPU consumption and performance remains roughly the same and therefore stable in the cluster machines as can be appreciated.

We have also taken measurements for the following workloads obtaining results as follows:

\paragraph{Workloads for YCSB}

We have tested our implementation in HBase with several built-in workloads from YCSB plus one custom workload with 100\% writes to stress the database intensively, because target updates in social networks as previously mentioned, are mostly all about changes and new insertions.

Figure~\ref{fig-bandwidth-worloada} shows three different sets of Qualities of Data for the same workload (A):
\begin{enumerate}
\item{YCSB workload A (R/W - 50/50)}
	\begin{itemize}
		\item No QoD enforced.
		\item QoD fulfillment of $\sigma$=0.5\% of total updates to be replicated.
		\item QoD fulfillment of $\sigma$=2\% of total updates to be replicated.
	\end{itemize}

During the execution of the workload A, in Figure~\ref{fig-bandwidth-worloada}, the highest peaks in replication traffic are observed without any type of QoD, i.e. just using plain HBase. This is due to the nature of eventual consistency itself and the buffering mechanisms in HBase.

With a QoD enabled as shown in the other two graphs, we rather control traffic of updates from being unbounded to a limited size, and accordingly saving resources' utilization, while suiting applications that require smaller amounts of updates as they only propagated as a group, when they are just needed.

We observe that a higher QoD implies replication traffic less often, although interactions reach high values on Bytes as they need to send more data. Small QoD optimizes the usage of resources while sending priority updates more frequently (this could be the case of wall posts in a social network).

\begin{figure*}[!t]
\centering
\includegraphics[width=0.7\linewidth]{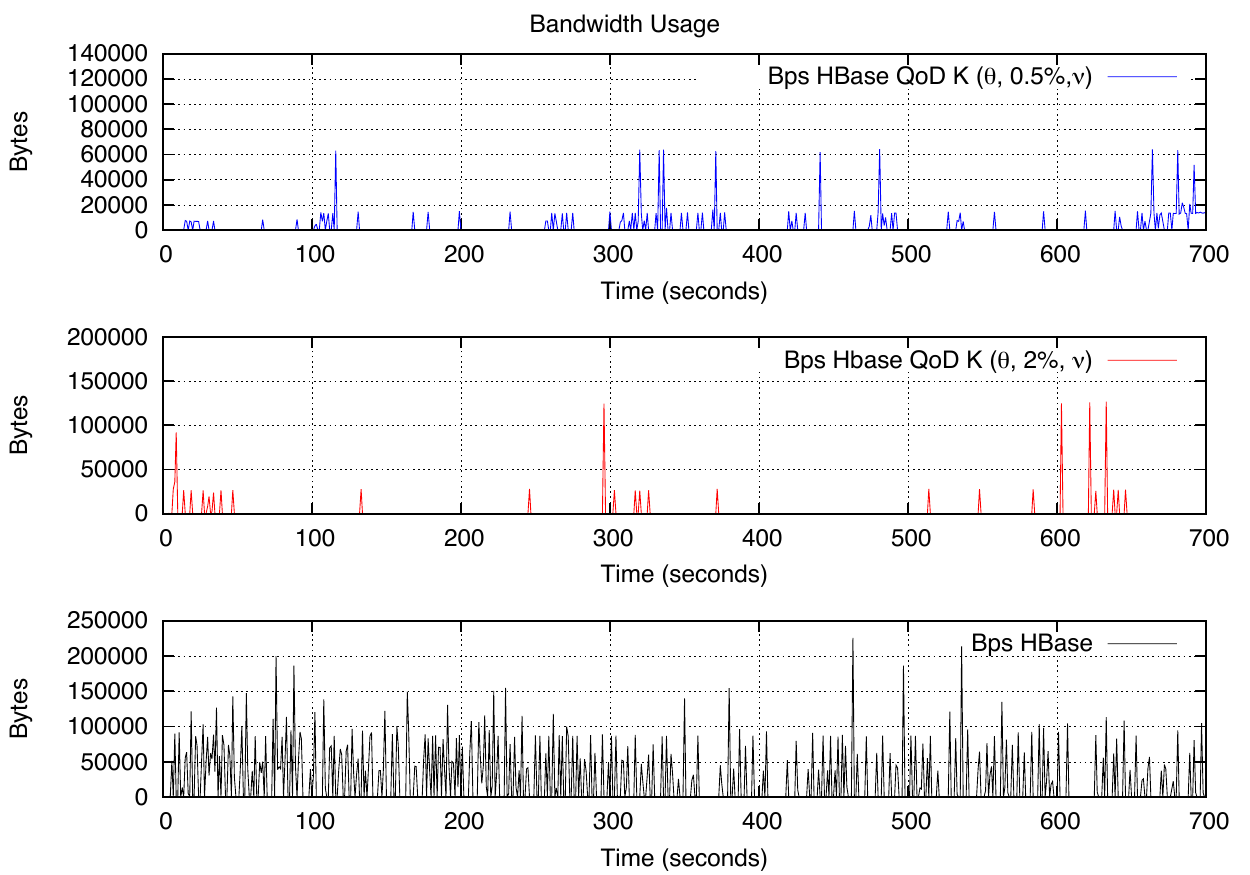}
\caption{Bandwidth usage for Workload A with zipfian distribution, using 5M records using QoD bounds of 0.5 and 2\% in the $\sigma$ of K.}
\label{fig-bandwidth-worloada}
\end{figure*}

\item{YCSB workload A modified (R/W - 0/100)}
	\begin{itemize}
		\item No QoD enforced.
		\item QoD fulfillment of $\sigma$=0.5\% of total updates to be replicated.	
		\item QoD QoD fulfillment of $\sigma$=2\% of total updates to be replicated.  
	\end{itemize}
\end{enumerate}

\begin{figure*}[!t]
\centering
\includegraphics[width=0.7\linewidth]{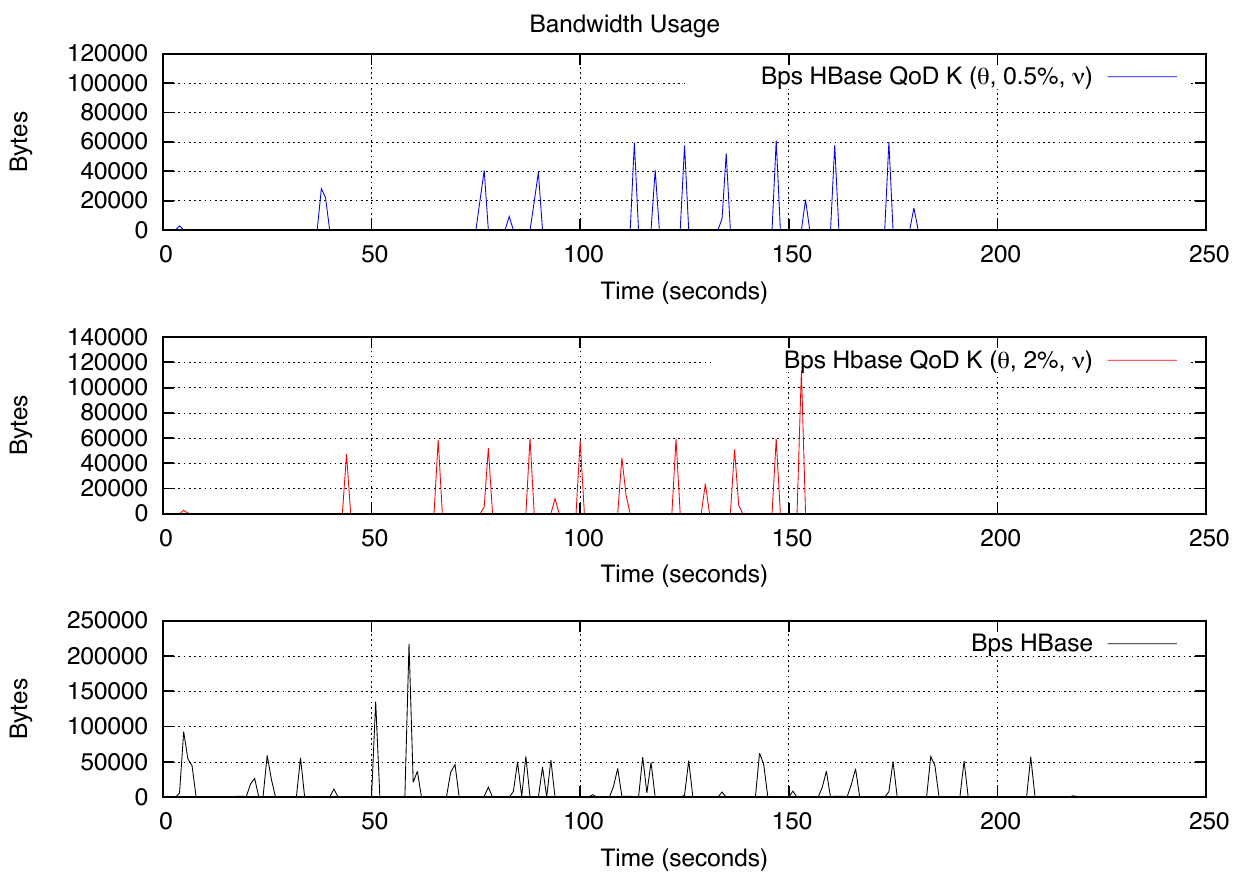}
\caption{Bandwidth usage for custom Workload with uniform distribution, using 5M records and QoD bounds of 0.5 and 2\% in the $\sigma$ of K.}
\label{fig-bandwidth-worloada-modified}
\end{figure*}

In Figure ~\ref{fig-bandwidth-worloada-modified} we can see how a write intensive workload performs using a QoD. Similar results are expected and later also confirmed in this graph (please note the scale of the Y axis is modified in order to show the relevant difference in Bytes more accurately).
  For smaller QoD (0.5\%), overall we see lower peaks in bandwidth usage than with plain HBase, as well as in the following measurement used for QoD 2.0\% (having that one higher maximum peak values than the previous QoD). Finally HBase with no modifications shows a much larger number of Bytes when coming to maximum bandwidth consumption.
  Note we are not measuring, or find relevant, in any of these scenarios, to realize savings on average bandwidth usage. The principal source of motivation of the paper is to find a way of controlling the usage of the resources in a data center. Also, to be able to leverage the trading of strong for eventual consistency with a more robust atomic grouping of operations using vector bounded data semantics.

%% file: conclusion.tex
Performance in HBase improves as the number of resources increases, for instance with more memory available~\cite{Carstoiu:2010}, but it is not always trivial to scale further following that approach. Therefore, having ways of providing different levels of consistency to users regarding data in cloud environments translates into substantial traffic savings and potentially associated costs to service providers. That is a relevant matter already discussed for consistency cost-efficiency~\cite{chihoub:2013}.

In this paper we presented HBase-QoD, a module that uses quality-of-data (QoD) to envision the goal of a tunable consistency model for geo-replication. The framework allows bounds on data to be used in order to perform selective replication in a more controlled and timely-fashion than usual eventually consistent approaches in these sort of data stores.

With the semantics presented we trade-off short timed consistency with wide area bandwidth cost savings during peak loads. Achieving the last, can help to significantly reduce replication overhead between data centers when there are periods of disconnection or bottlenecks in the network. We evaluated our implementation on top of HBase clusters distributed across several locations showing relevant results for that.

In summary, a successful model based on vector-field divergence mechanisms~\cite{Santos:2010} was implemented and shows how HBase consistency can be tuned at the core-system level, without requiring intrusion to the data schema and avoiding more middle-ware overhead such as in~\cite{Das:2010}. In our case, experimental results indicate that we are able to maintain an acceptable throughput, reduce latency peaks, as well as optimize bandwidth usage. In the future we would like to conduct more experiments using Amazon EC2 infrastructure and also several other cluster locations in partner-universities if there is any chance to do so. 

%% file: alvaro-hbase-qod-cloudcom2013.bbl
\begin{thebibliography}{10}
\providecommand{\url}[1]{#1}
\csname url@samestyle\endcsname
\providecommand{\newblock}{\relax}
\providecommand{\bibinfo}[2]{#2}
\providecommand{\BIBentrySTDinterwordspacing}{\spaceskip=0pt\relax}
\providecommand{\BIBentryALTinterwordstretchfactor}{4}
\providecommand{\BIBentryALTinterwordspacing}{\spaceskip=\fontdimen2\font plus
\BIBentryALTinterwordstretchfactor\fontdimen3\font minus
  \fontdimen4\font\relax}
\providecommand{\BIBforeignlanguage}[2]{{%
\expandafter\ifx\csname l@#1\endcsname\relax
\typeout{** WARNING: IEEEtran.bst: No hyphenation pattern has been}%
\typeout{** loaded for the language `#1'. Using the pattern for}%
\typeout{** the default language instead.}%
\else
\language=\csname l@#1\endcsname
\fi
#2}}
\providecommand{\BIBdecl}{\relax}
\BIBdecl

\bibitem{Brewer:2002}
\emph{Brewer's Conjecture and the Feasibility of Consistent Available
  Partition-Tolerant Web Services.}, 2002.

\bibitem{Li:2012}
\BIBentryALTinterwordspacing
C.~Li, D.~Porto, A.~Clement, J.~Gehrke, N.~Pregui\c{c}a, and R.~Rodrigues,
  ``Making geo-replicated systems fast as possible, consistent when
  necessary,'' in \emph{Proceedings of the 10th USENIX conference on Operating
  Systems Design and Implementation}, ser. OSDI'12.\hskip 1em plus 0.5em minus
  0.4em\relax Berkeley, CA, USA: USENIX Association, 2012, pp. 265--278.
  [Online]. Available: \url{http://dl.acm.org/citation.cfm?id=2387880.2387906}
\BIBentrySTDinterwordspacing

\bibitem{Saphiro:2011}
\BIBentryALTinterwordspacing
N.~P. Marc~Shapiro and M.~Z. Carlos~Baquero, ``Conflict-free replicated data
  types,'' INRIA, rocq, rr RR-7687, July 2011. [Online]. Available:
  \url{http://lip6.fr/Marc.Shapiro/papers/RR-7687.pdf}
\BIBentrySTDinterwordspacing

\bibitem{Vfc3:2012}
\BIBentryALTinterwordspacing
S.~Esteves, J.~a. Silva, and L.~Veiga, ``Quality-of-service for consistency of
  data geo-replication in cloud computing,'' pp. 285--297, 2012. [Online].
  Available: \url{http://dx.doi.org/10.1007/978-3-642-32820-6\_29}
\BIBentrySTDinterwordspacing

\bibitem{Burckhardt:2012}
\BIBentryALTinterwordspacing
S.~Burckhardt, D.~Leijen, M.~Fahndrich, and M.~Sagiv, ``Eventually consistent
  transactions,'' in \emph{Proceedings of the 21st European conference on
  Programming Languages and Systems}, ser. ESOP'12.\hskip 1em plus 0.5em minus
  0.4em\relax Berlin, Heidelberg: Springer-Verlag, 2012, pp. 67--86. [Online].
  Available: \url{http://dx.doi.org/10.1007/978-3-642-28869-2\_4}
\BIBentrySTDinterwordspacing

\bibitem{Carstoiu:2010}
D.~Carstoiu, A.~Cernian, and A.~Olteanu, ``Hadoop hbase-0.20.2 performance
  evaluation,'' in \emph{New Trends in Information Science and Service Science
  (NISS), 2010 4th International Conference on}, May 2010, pp. 84 --87.

\bibitem{Sovran:2011}
\BIBentryALTinterwordspacing
Y.~Sovran, R.~Power, M.~K. Aguilera, and J.~Li, ``Transactional storage for
  geo-replicated systems,'' in \emph{Proceedings of the Twenty-Third ACM
  Symposium on Operating Systems Principles}, ser. SOSP '11.\hskip 1em plus
  0.5em minus 0.4em\relax New York, NY, USA: ACM, 2011, pp. 385--400. [Online].
  Available: \url{http://doi.acm.org/10.1145/2043556.2043592}
\BIBentrySTDinterwordspacing

\bibitem{Duke:2001}
\BIBentryALTinterwordspacing
``Combining generality and practicality in a conit-based continuous consistency
  model for wide-area replication,'' Washington, DC, USA, pp. 429--, 2001.
  [Online]. Available: \url{http://dl.acm.org/citation.cfm?id=876878.879318}
\BIBentrySTDinterwordspacing

\bibitem{Santos:2010}
\BIBentryALTinterwordspacing
L.~Veiga, A.~P. Negr{\~a}o, N.~Santos, and P.~Ferreira, ``Unifying divergence
  bounding and locality awareness in replicated systems with vector-field
  consistency,'' \emph{J. Internet Services and Applications}, vol.~1, no.~2,
  pp. 95--115, 2010. [Online]. Available:
  \url{http://dx.doi.org/10.1007/s13174-010-0011-x}
\BIBentrySTDinterwordspacing

\bibitem{Chang:2006}
\BIBentryALTinterwordspacing
F.~Chang, J.~Dean, S.~Ghemawat, W.~C. Hsieh, D.~A. Wallach, M.~Burrows,
  T.~Chandra, A.~Fikes, and R.~E. Gruber, ``Bigtable: a distributed storage
  system for structured data,'' in \emph{Proceedings of the 7th USENIX
  Symposium on Operating Systems Design and Implementation - Volume 7}, ser.
  OSDI '06.\hskip 1em plus 0.5em minus 0.4em\relax Berkeley, CA, USA: USENIX
  Association, 2006, pp. 15--15. [Online]. Available:
  \url{http://dl.acm.org/citation.cfm?id=1267308.1267323}
\BIBentrySTDinterwordspacing

\bibitem{Lakshman:2010}
\BIBentryALTinterwordspacing
A.~Lakshman and P.~Malik, ``Cassandra: a decentralized structured storage
  system,'' \emph{SIGOPS Oper. Syst. Rev.}, vol.~44, no.~2, pp. 35--40, Apr.
  2010. [Online]. Available: \url{http://doi.acm.org/10.1145/1773912.1773922}
\BIBentrySTDinterwordspacing

\bibitem{Corbett:2012}
\BIBentryALTinterwordspacing
J.~C. Corbett, J.~Dean, M.~Epstein, A.~Fikes, C.~Frost, J.~J. Furman,
  S.~Ghemawat, A.~Gubarev, C.~Heiser, P.~Hochschild, W.~Hsieh, S.~Kanthak,
  E.~Kogan, H.~Li, A.~Lloyd, S.~Melnik, D.~Mwaura, D.~Nagle, S.~Quinlan,
  R.~Rao, L.~Rolig, Y.~Saito, M.~Szymaniak, C.~Taylor, R.~Wang, and
  D.~Woodford, ``Spanner: Google's globally-distributed database,'' in
  \emph{Proceedings of the 10th USENIX conference on Operating Systems Design
  and Implementation}, ser. OSDI'12.\hskip 1em plus 0.5em minus 0.4em\relax
  Berkeley, CA, USA: USENIX Association, 2012, pp. 251--264. [Online].
  Available: \url{http://dl.acm.org/citation.cfm?id=2387880.2387905}
\BIBentrySTDinterwordspacing

\bibitem{Lloyd:2011}
\BIBentryALTinterwordspacing
W.~Lloyd, M.~J. Freedman, M.~Kaminsky, and D.~G. Andersen, ``Don't settle for
  eventual: scalable causal consistency for wide-area storage with cops,'' in
  \emph{Proceedings of the Twenty-Third ACM Symposium on Operating Systems
  Principles}, ser. SOSP '11.\hskip 1em plus 0.5em minus 0.4em\relax New York,
  NY, USA: ACM, 2011, pp. 401--416. [Online]. Available:
  \url{http://doi.acm.org/10.1145/2043556.2043593}
\BIBentrySTDinterwordspacing

\bibitem{Yu:2000}
\BIBentryALTinterwordspacing
H.~Yu and A.~Vahdat, ``Design and evaluation of a continuous consistency model
  for replicated services,'' in \emph{Proceedings of the 4th conference on
  Symposium on Operating System Design \& Implementation - Volume 4}, ser.
  OSDI'00.\hskip 1em plus 0.5em minus 0.4em\relax Berkeley, CA, USA: USENIX
  Association, 2000, pp. 21--21. [Online]. Available:
  \url{http://dl.acm.org/citation.cfm?id=1251229.1251250}
\BIBentrySTDinterwordspacing

\bibitem{Cooper:2008}
\BIBentryALTinterwordspacing
B.~F. Cooper, R.~Ramakrishnan, U.~Srivastava, A.~Silberstein, P.~Bohannon,
  H.-A. Jacobsen, N.~Puz, D.~Weaver, and R.~Yerneni, ``Pnuts: Yahoo!'s hosted
  data serving platform,'' \emph{Proc. VLDB Endow.}, vol.~1, no.~2, pp.
  1277--1288, Aug. 2008. [Online]. Available:
  \url{http://dl.acm.org/citation.cfm?id=1454159.1454167}
\BIBentrySTDinterwordspacing

\bibitem{chihoub:2013}
\BIBentryALTinterwordspacing
H.-E. Chihoub, S.~Ibrahim, G.~Antoniu, and M.~P{\'e}rez,
  ``\BIBforeignlanguage{Anglais}{{Consistency in the Cloud:When Money Does
  Matter!}}'' in \emph{\BIBforeignlanguage{Anglais}{{CCGRID 2013- 13th IEEE/ACM
  International Symposium on Cluster, Cloud and Grid Computing}}}, Delft,
  Pays-Bas, May 2013. [Online]. Available:
  \url{http://hal.inria.fr/hal-00789013}
\BIBentrySTDinterwordspacing

\bibitem{Das:2010}
\BIBentryALTinterwordspacing
S.~Das, D.~Agrawal, and A.~El~Abbadi, ``G-store: a scalable data store for
  transactional multi key access in the cloud,'' in \emph{Proceedings of the
  1st ACM symposium on Cloud computing}, ser. SoCC '10.\hskip 1em plus 0.5em
  minus 0.4em\relax New York, NY, USA: ACM, 2010, pp. 163--174. [Online].
  Available: \url{http://doi.acm.org/10.1145/1807128.1807157}
\BIBentrySTDinterwordspacing

\end{thebibliography}
